\documentclass{raa}

\usepackage{graphicx,times}             

\begin{document}

\title{Linear nonadiabatic pulsation models of the double-mode Cepheid V371 Per}

   \volnopage{Vol.0 (200x) No.0, 000--000}      
   \setcounter{page}{1}          

\author{Toshihito Ishida \inst{} }
\institute{Center for Astronomy, Institute of Natural and Environmental Science, University of Hyogo, Sayo-cho, Hyogo, 679-5313, Japan; 
 {\it ishida@nhao.jp} }

\date{Received~~2009 month day; accepted~~2009~~month day}

\label{firstpage}

\abstract{ 
Recently found double-mode Cepheid with the shortest period in our galaxy 
and abnormal period ratio, V371 Per is investigated by 
linear nonadiabatic pulsation models.  
V371 Per is likely to be crossing the instability strip 
for the first time, 
because mass derived from pulsation models is larger 
than the evolution mass for the second or higher crossing objects.  
This result seems to support the conclusion obtained from the 
spectroscopic observation.  
We also found that models with observed period and period ratio of V371 Per 
need to have mass and Teff in a narrow range which shifts as 
heavy element abundance $Z$ changes.  
We have checked the agreement between Teff ranges estimated observationally 
and derived from pulsation models using observational $Z$.  
We found that those ranges overlap marginally.  
We need more spectroscopic estimations of the Teff and $[Fe/H]$, 
and more photometric monitoring to estimate the evolutionally 
period change for confirmation of our result.  
}

\authorrunning{T. Ishida }            
\titlerunning{Linear Nonadibatic Pulsation Models of the Double-Mode Cepheid V371 Per}  

\maketitle

\section{Introduction}
Double-mode cepheids are the subgroup of the classical Cepheids, 
which pulsate in two radial pulsation modes simultaneously (For a review 
article, see for example Petersen \&\ Takeuti~(\cite{Pe01}) and references there in).  
These stars are checked for the difference from the single-mode Cepheids 
for many aspects, however no difference is reported up to now.  
We can interpret that double-mode cepheids are the normal classical Cepheids, 
which happen to pulsate in two pulsation modes.  
There are 22 clearly established galactic members of this subgroup.  
The period ratios of the two periods are from $0.6967$ to $0.7195$ 
with two exceptions with ratios around $0.80$ (CO Aur and HD304373), 
which are considered as the first overtone and the second overtone pulsators.  
The periods of the fundamental mode for other 20 stars are 
between $2.139$ and $6.293$ days.  

Recently, V371 Per is reported as a new candidate member 
of this subgroup (Wils et al.~\cite{Wi10}).  
V371 Per was classified as irregular variable by Weber(\cite{We64}), 
but Schmidt et al.(\cite{Sc95}) suggested this star to be a double-mode cepheid from 
the detected period and excessive scatter in the light curve.  
Thus, Wils et al.(\cite{Wi10}) made new observations and indicated that 
this star is indeed a double-mode Cepheids with the shortest 
fundamental mode period of 1.738 days and unusually 
high period ratio of 0.731.  They discussed that 
the high value of the period ratio may indicate that 
the metalicity of this star is much lower than the other 
Galactic double-mode cepheids from the relation between 
period and period ratio versus metalicity Z discussed by Buchler \&\ Szabo(\cite{Bu07}). 
They also pointed out that this star is in the Thick Disk 
or the Halo of our Galaxy.  
Furthermore, Kovtyukh et al.(\cite{Ko12, Ko16}) discussed that 
V371 Per can be considered as a first crossing object 
from its peculiar abundances, especially from richness of the lithium.  

Wils et al.(\cite{Wi10}) further discussed on some properties of this star, 
however, the properties they deduced from pulsation models 
are not convincing because of the lack of computations with 
appropriate parameters due to its short period and other unusual 
properties.  We would like to present here some results 
of linear nonadiabatic pulsation models with special interest to 
this unusual star.  

\section{Models}

For calculating linear non-adiabatic period, we used the same code used 
in Ishida(\cite{Is95}), namely, Castor(\cite{Ca71}) type procedure with OPAL opacity
tables (Iglesias, et al.~\cite{Ig92}; Rogers \&\ Iglesias~\cite{Ro92}).
The effect of convection seems not efficient for our purpose, 
as the derived metalicity for V367 Sct and Y Car 
using linear non-adiabatic calculations in Ishida(\cite{Is95}) 
is not far from recent observation (Andrievsky et al.~\cite{An02}).  
Therefore, the effect of convective energy transport is not included 
in all of the models.  

The luminosity of the model is estimated from the 
Fouqu\'e et al.~(\cite{Fo07})'s period-luminosity relation for Galactic Cepheids 
and Torres(\cite{To10})'s polynomial fits for the Flower(\cite{Fl96})'s bolometric correction.  
Luminotity of the V371 Per is estimated to be 460L\sun. 
Considering differences between distance derived from 
the period--luminosity relations in the different photometric bands 
into account, $L/L\sun=500$ and $420$ models are also 
examined.  

The effective temperature of V371 Per is reported as; 
6000K or lower from unreddened $B-V$ by Wils et al.(\cite{Wi10}), 
6215K from average of low-resolution spectra by Schmidt(\cite{Sc11}), 
and 5950 $\sim$ 6213K from high resolution spectra by Kovtyukh et al.(\cite{Ko12}). 
Considering spectral classification as G0 by Bond(\cite{Bo78}) into account, 
we set wide parameter range for effective temperature 
from 7000 to 5400K with 100K difference. 

Observational estimates of the iron abundance are; 
$[Fe/H]= -0.05$ to $-0.40$ from low-resolution spectra by Schmidt(\cite{Sc11}), 
and $[Fe/H]= -0.42$ from high-resolution spectra by Kovtyukh et al.(\cite{Ko12}). 
Thus, chemical compositions of the calculated models are 
$Z = 0.001, 0.004, 0.01, 0.02$ and $0.03$. 

The mass of the models are changed from 1.5 to 6.0 $M\sun$ 
with 0.5$M\sun$ difference.  
Selected parameter range is summarized in Table~\ref{tbl1}. 

\begin{table}
\begin{center}
\caption{Summary of the Model Parameters.}
\label{tbl1}
\begin{tabular}{@{}lccccc}
\hline
Parameters & \\
\hline
Luminosity (in L\sun) & \multicolumn{5}{c}{420, 460, 500} \\
Teff & \multicolumn{5}{c}{7000 $\sim$ 5400} \\
X & \multicolumn{5}{c}{0.70} \\
Z & 0.001 & 0.004 & 0.01 & 0.02 & 0.03 \\ 
Mass (in M\sun) & \multicolumn{5}{c}{1.5 $\sim$ 6.0} \\
\hline
\end{tabular}
\end{center}
\end{table}

\section{Results of the linear nonadiabatic models}
\begin{figure}
\centering
\includegraphics[width=130mm]{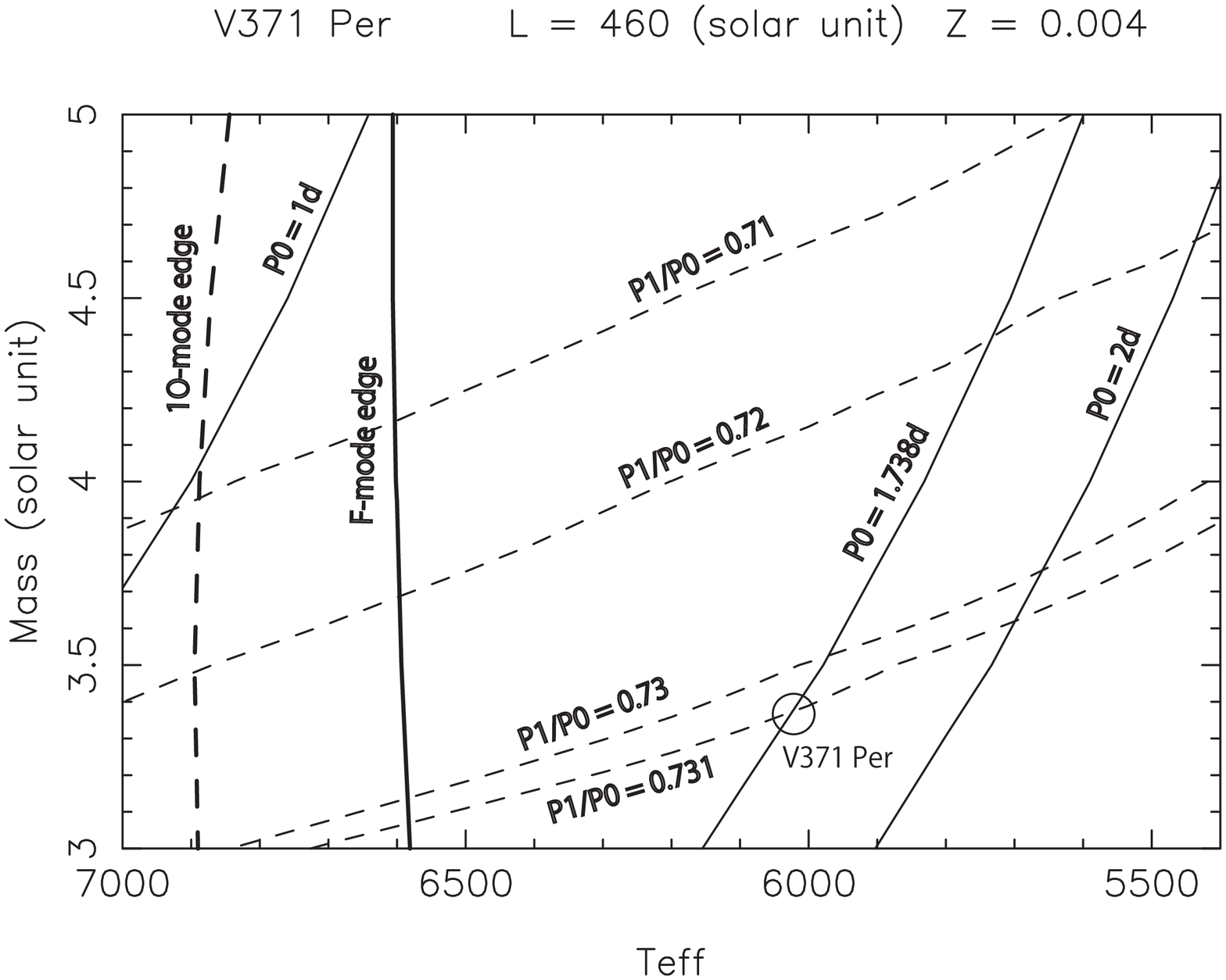}
\caption{Contour plot of the fundamental period, period ratio, 
and blue edges on Teff vs Mass plane for $L=460L\sun$ and $Z=0.004$ models.  
Abscissa is the effective temperature and ordinate is the mass 
in solar unit. 
Solid lines are fundamental period 
contour with corresponding period value on the side of each line.  
Dashed lines are period ratio contour. Notifications are the same as the 
fundamental period. 
Thick lines are instability blue edge for the fundamental(solid) and 
the first overtone(dashed), respectively.  
The position of the model parameter satisfies all of the criteria, 
namely, the observed period, the observed period ratio, and 
both of the fundamental and the first overtone modes are 
pulsationally unstable, is indicated by an open circle.  }
\label{fig1}
\end{figure}

As a typical example, the results for $L = 460L\sun$, $Z=0.004$ 
are summarized in Fig.~\ref{fig1}.  
The observed period 1.737 days and period ratio 0.731 is 
realized around $M=3.35M\sun$ and Teff=6020.  
In $L=420L\sun$ models, both of the instability edges move 
slightly to the higher-temperature direction, 
the constant period lines move to the lower-temperature 
direction, and the constant period ratio lines move to 
the lower-mass direction.  Thus the observed 
period and period ratio are realized at the 
lower-mass and lower-temperature point than $L=460L\sun$ model.  
In $L=500L\sun$ models, changes are oposite directions and 
the observed period and period ratio are realized at the 
higher-mass and higher-temperature point than 
$L=460\sun$ models.  

\begin{figure}
\centering
\includegraphics[width=130mm]{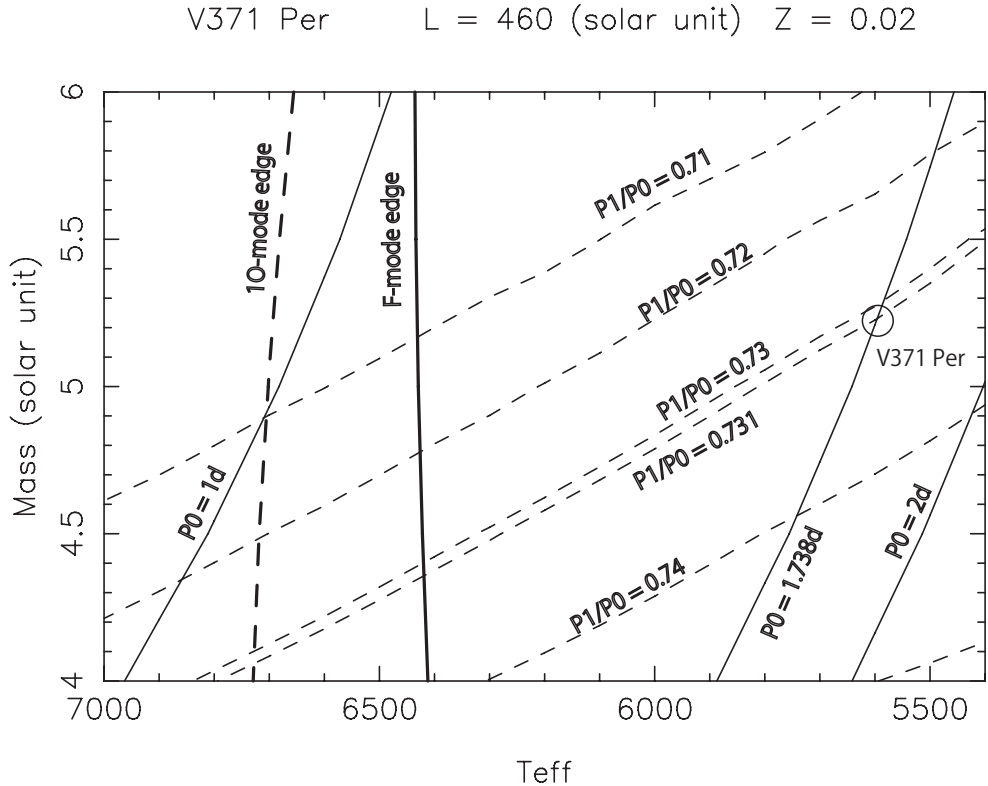}
\caption{Contour plot of the fundamental period, period ratio, 
and blue edges on Teff vs Mass plane for $L=460\sun$ and $Z=0.02$ models.  
Notations are the same as Fig~\ref{fig1}.  }
\label{fig2}
\end{figure}

The results for the $L=460L\sun$ and $Z=0.02$ models are shown in 
Fig.~\ref{fig2}.  
For comparison with Fig.~\ref{fig1}, we need to remark that 
the presented mass range is changed from 1.5$\sim$5.0 $M\sun$ for $Z=0.004$ models 
to 4.0$\sim$6.0 $M\sun$ for $Z=0.02$ models.  
In the $Z=0.02$ models, both of the instability blue edges move 
to the lower temperature direction, 
the constant period lines move to the higher-temperature 
direction, and the constant period ratio lines 
move to the higher-mass direction, comparing to the 
$Z=0.004$ models.  
It is also worth noting that the slope of the 
constant period ratio lines is steeper in the $Z=0.02$ models.  
Thus the observed period and period ratio is realized 
at the higher-mass and lower-temperature point than the 
$Z=0.004$ model, namely, around $M=5.22M\sun$ and Teff=5600.  

\begin{figure}
\centering
\includegraphics[width=130mm]{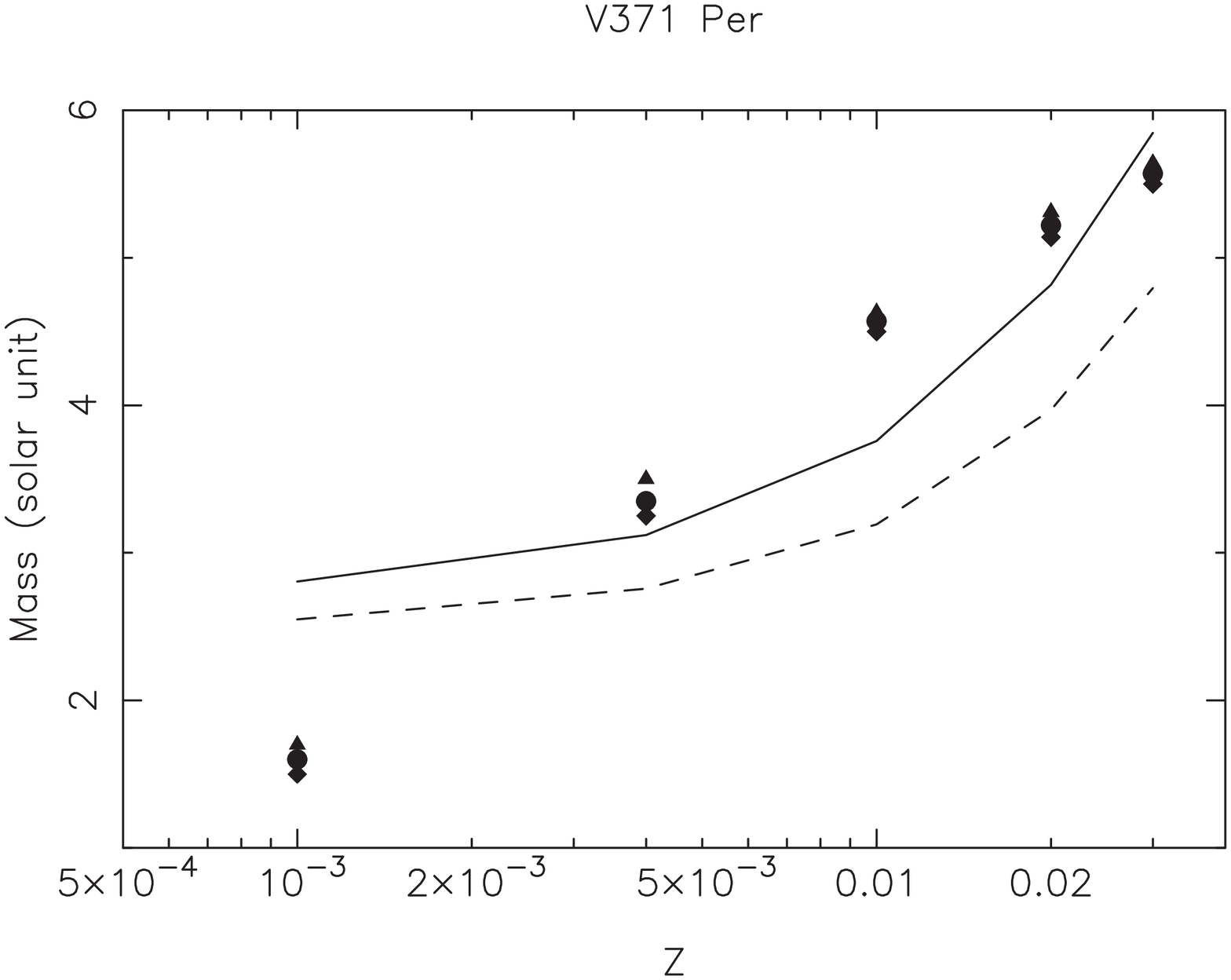}
\caption{Derived model mass to realize the observed period and 
period ratio. Abscissa is the heavy element abundance Z of the models. 
Ordinate is the derived mass in solar unit.  Results from the 
$L=460L\sun$ models are plotted by dots. $L=500L\sun$ and $L=420L\sun$ models 
are also plotted by filled triangles, and filled squares, respectively. 
Masses estimated from Becker, et al.(1977)'s mass-luminosity relations for the first 
crossing are shown with the solid line, and those for the second crossing 
are with the dashed line. }
\label{fig3}
\end{figure}

\begin{figure}
\centering
\includegraphics[width=130mm]{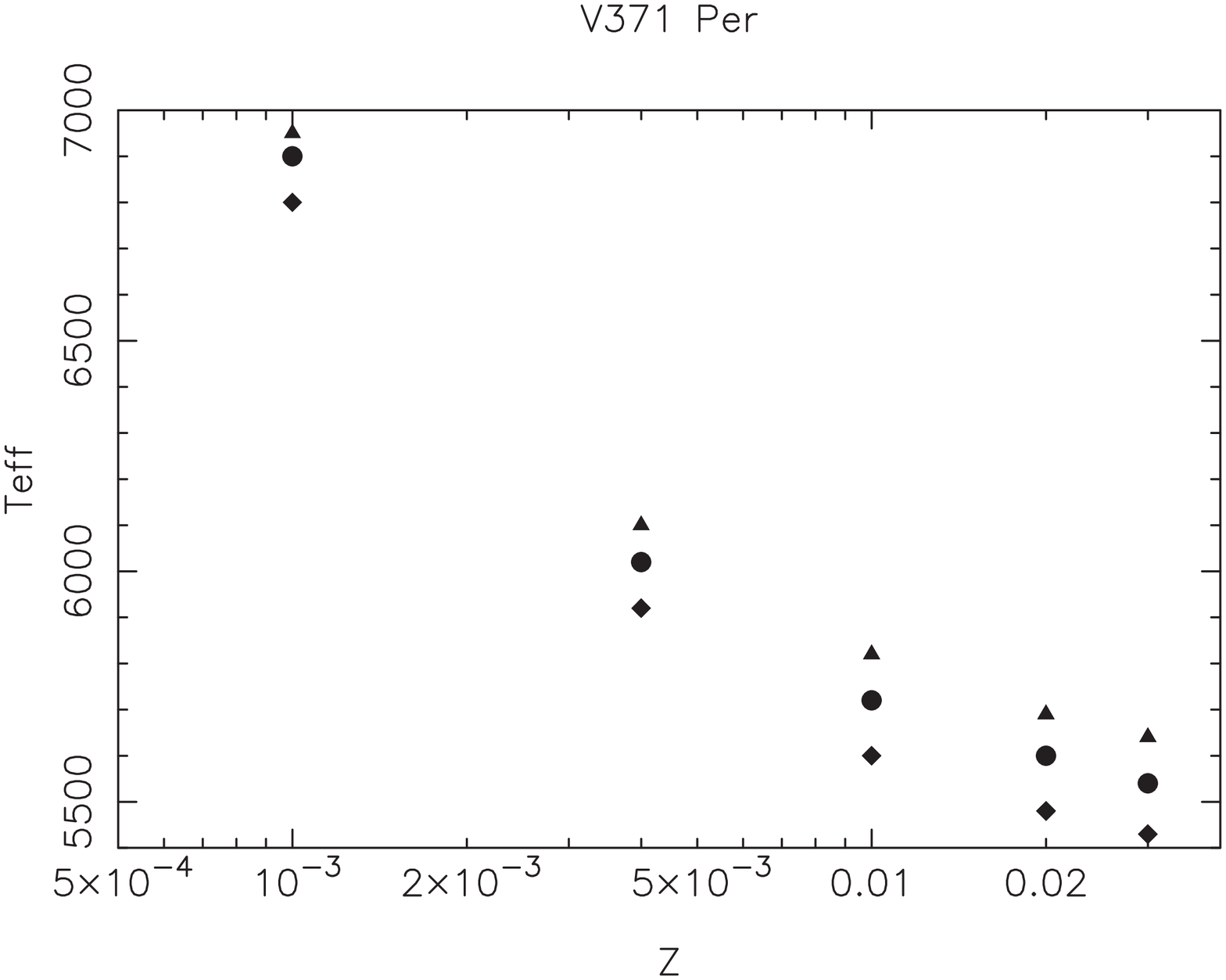}
\caption{Derived model effecive temperature to realize the observed 
period and period ratio.  Abscissa is the same as Fig~\ref{fig3}. 
Ordinate is the effective temperature.  Marks used in this plot are 
the same as Fig~\ref{fig3}. }
\label{fig4}
\end{figure}

Derived mass and effective temperature to realize the observed 
period and period ratio is summarized in Fig.~\ref{fig3} and Fig.~\ref{fig4}. 
As pointed out in the description of the Fig.~\ref{fig2}, 
the Mass of the satisfactory model increases as Z increases, 
and the Teff of that decreases.  
We need to remark that both of the fundamental and the first overtone modes 
are pulsationally stable in the $Z=0.001$ model.  

\section{Discussion}
As shown in Figs.\ref{fig3} and \ref{fig4}, the observed period and 
period ratio is realized with wide range of the chemical composition, 
however, the mass and effective temperature needs to change as chemical composition.  
Namely, the model with observed period and period ratio 
should be in a narrow band on the mass and Teff versus Z planes.  
Therefore, when we fix Z to the observationally estimated value, 
whether the observationally estimated Teff is in the range 
derived from the pulsation model or not will be a check 
of the agreement between observation and the pulsation theory.  
From 4 high-resolution spectra, Kovtyukh et al.~(\cite{Ko12}) derived 
$[Fe/H] = -0.42$ and Teff = 5950 $\sim$ 6213.  
The simple conversion from $[Fe/H]$ to $Z$ indicates $Z\sim 0.007$.  
From Fig.\ref{fig4}, we can estimate Teff = 5700 $\sim$ 5900.   
Considering above observed Teff range is from only 4 spectra into account, 
we can consider that the observations and our estimation from pulsation models 
is marginally overlapping.  
More observational estimations of the Teff and chemical composition 
are needed to confirm whether our results agrees with the 
observations.  

In the low Z models, helium content $Y$ used here is especially high.  
Therefore, additional calculation for $(X, Z) =(0.80, 0.01)$ models are performed 
to confirm the effect of high $Y$ values.  
The process to search for the mass and Teff to satisfy the observed properties of V371 Per 
is the same as that used in the previous section.  
The obtained mass is 0.2M$\sun$ larger than that obtained for $(X, Z)=(0.70, 0.01)$ models. 
The obtained Teff is almost unchanged.  Thus, the result presented in the previous section will 
not be largely affected by using high $Y$ values.  

The effect of inclusion of convection on the fundamental period 
and period ratio of the first overtone mode to the fundamental mode 
is investigated by Saio et al.~(\cite{Sa77}).  Their result indicate that 
the fundamental period becomes slightly longer and the period 
ratio becomes smaller as the mixing length $l/H_p$ becomes larger.  
On the Fig.\ref{fig1} and \ref{fig2}, inclusion of the convection 
will shift the period constant lines slightly to the higher-mass direction 
and shift the period ratio constant lines to the lower-mass direction.  
Because the effect on the period ratio is larger than the effect 
on the period, the position of the model with observed  period 
and period ratio will shitf to the lower-mass direction, namely, 
the derived mass will be smaller and derived effective 
temperature will be higher.  
Therefore, inclusion of convection will make better agreement 
between the observation and our estimation from pulsation models for Teff 
discussed in the previous paragraph.  

In Fig.\ref{fig3}, the mass for the calculated luminosity 
derived from the evolutional relations 
by Becker et al.~(\cite{Be77}) is also plotted.  The masses estimated by the 
mass-luminosity relation for the second crossing objects are 
shown by the dashed line and those for the first crossing 
objects are shown by the solid line.  
Our results indicate that the pulsationally derived mass is 
massive than the second crossing mass in almost all range of Z.  
It is even massive than the first crossing mass from about 
$Z= 0.004$ to $0.02$.  At the observed chemical composition $Z \sim 0.007$, 
the mass estimated from our results is $\sim 4.1M\sun$. 
Considering the effect of convection 
discussed in the previous paragraph into account, 
the excess over the first crossing mass will be reduced.  
Thus, our result from pulsation models seems to indicate that 
V371 Per is too massive as the second or higher crossing object, 
and is likely to be the first crossing object as disscussed by 
Kovtyukh et al.~(\cite{Ko12, Ko16}) from spectroscopic observations.  

More modern evolution calculation with OPAL opacity 
is presented by Bono et al.~(\cite{Bo00}) and they obtained a mass-luminosity 
relation for classical Cepheids.  
Although, their relation is for the second and the third 
crossing objects, and for the objects in the mass 
range $4$ -- $15 M\sun$, the estimated mass for $L=460L\sun$ 
with $(Y, Z) = (0.293, 0.007)$ is about $3.34M\sun$ from their relation. 
This estimation is less massive than the mass estimated from 
our results in the previous paragraph, and seems to 
support that V371 Per is a first crossing object. 
It is also worth noting that the $4.0M\sun$ model in Bono et al.~(\cite{Bo00})'s 
table 7 with $(Y, Z) = (0.23, 0.004)$ has similar first crossing 
luminosity $\log L/L\sun = 2.620$ ($416L\sun$) 
to the present model $460L\sun$ and 
with shorter period $\log P = 0.0171$ (1.040 days) 
at higher temperature $\log T = 3.808$ (6426K).  
Our consequnce that V371 Per is likely to be a first crossing 
object seems also valid with state-of-the-art evolution calculation.  

Wils et al.~(\cite{Wi10}) reported that both of the fundamental and the first 
overtone modes frequencies are decreased. 
Frequency decrease means period increase.  
Period increase is realized when the Cepheid is the 1st, 3rd, or 5th 
crossing of the instability strip.  
Effects of the chemical composition and the periods on the 
evolutionary period change of Cepheids is already 
investigated by Saitou(\cite{Sa89}) using Becker et al.(\cite{Be77})'s evolution calculation.  
His results indidate that the evolutionary period changes 
for the 1st, 4th+5th crossing is larger than those for 
the 2nd and 3rd crossing, especially at the short period region.  
He also derived abundance dependent period change--period relations for 
each crossing.  We can estimate the period change for V371 Per 
to be about $\log[\Delta P/P]_{100} = -3.11$ from Wils et al.~(\cite{Wi10})'s Figure 5.  
On the other hand, we can estimate from the Saitou(\cite{Sa89})'s fittings, 
that $\log[\Delta P/P]_{100} \sim -2.88$ for the first crossing and 
$\sim -4.67$ for the 3rd crossing with $Z = 0.007$.  
Consequently, the observed period change seems to fit to the first crossing.  
Recently, evolutionary period change is also discussed by Pietrukowics(\cite{Pi03}) and 
Turner et al.(\cite{Tu06}).  We can convert above period change to $\log {\dot P} \sim 0.063$ 
and ${\dot P}/P^2 \sim 12.14$.  Both values put V371 Per in the 
first crossing region of Fig.2 of Pietrukowics(\cite{Pi03}) and Fig.3 of Turner et al.(\cite{Tu06}). 
Therefore, the period change reported by Wils et al.(\cite{Wi10}) seems to 
support our result that V371 Per is a first crossing object 
derived from linear pulsation models.  
However, Wils et al.(\cite{Wi10}) reported that 
the period change of the V371 Per should be treated with caution, 
because this change highly depends on the photographic data. 
Thus, we need more photometric monitoring of 
this interesting and unusual Cepheid and more precise 
estimation of its period change.  

\section{Conclusions}
We have investigated recently found double-mode Cepheid V371 Per 
by linear nonadiabatic pulsation models.  
Our result indicates that V371 Per is likely to be a first crossing object, 
because the mass derived from linear pulsation results is larger 
than the evolutionary mass for the second crossing object.  
Thus, our result obtained from pulsation models seems to support 
the conclusion from spectroscopic observation by 
Kovtyukh et al.(\cite{Ko12, Ko16}). 
Decreases of the frequencies reported by Wils et al.(\cite{Wi10}) seems 
to support this consequence.  
We also found that models with observed period and period ratio of V371 Per 
need to have mass and Teff in a narrow range which shifts as 
heavy element abondance $Z$ changes.  
We have checked whether observational Teff is in the range 
derived from pulsation models using observational $Z$ or not.  
Two Teff ranges is marginally overlapping, 
and agreement between observation and our result seems to be marginal.  
We need more spectroscopic estimations of the Teff and $[Fe/H]$ 
to check our result, 
and more photometric monitoring to estimate the evolutionally 
period change and thus confirm evolutional status of 
V371 Per.  

\section*{Acknowledgments}
I would like to appreciate to my colleagues for their 
useful comments during this investigation.  
Computations are carried out mainly on the workstation prepared by 
the research grant of the University of the Hyogo.

\label{lastpage}

\end{document}